\documentclass[preprint]{elsarticle}

\usepackage[T1]{fontenc} % Use modern font encodings
 \usepackage{array}
 \usepackage{lmodern}\usepackage{mathpazo}\usepackage{microtype}
\usepackage{mathrsfs} \usepackage{amssymb} \usepackage{graphicx,color} \usepackage{bbm} \usepackage{multirow}
\usepackage{bm} \usepackage{mathrsfs} \usepackage{commath} \usepackage{stmaryrd}
\usepackage{color,soul}
\usepackage{amsmath}
\DeclareMathOperator\arctanh{arctanh}
 \usepackage{pdfpages} 

\let\om=\omega

\def\dd{\mathrm{d}}

\def\dd{\mathrm{d}}

\newcommand{\beq}{\begin{equation}} \newcommand{\eeq}{\end{equation}}

\newcommand{\argp}[1]{\left(#1\right)}

\def\mw{\omega_{\rm MW}}
\def\ome{\om_{\rm e}}
\def\omn{\om_{\rm n}}

\begin{document}

\begin{frontmatter}

\title{Unified understanding of the breakdown of thermal mixing dynamic nuclear polarization: the role of temperature and radical concentration}

\author[1,2]{Ludovica M. Epasto \fnref{fn1}}
\ead{ludovica.martina.epasto@univie.ac.at}
\author[3]{Thibaud Maimbourg \fnref{fn1}}
\ead{thibaud.maimbourg@ipht.fr}
\author[4]{Alberto Rosso \corref{cor1} }
\ead{alberto.rosso@u-psud.fr}
\author[1,2]{Dennis Kurzbach \corref{cor1} }
\ead{dennis.kurzbach@univie.ac.at}
\affiliation[1]{organization={University of Vienna, Faculty of Chemistry, Institute of Biological Chemistry}, addressline={Währinger Straße 38}, postcode={1090}, city={Vienna}, country={Austria}}
\affiliation[2]{organization={University of Vienna, Vienna Doctoral School in Chemistry (DoSChem)}, addressline={Währinger Str. 42}, postcode={1090}, city={Vienna}, country={Austria}}
\affiliation[3]{organization={Universit\'e Paris-Saclay, CNRS, CEA, Institut de physique th\'eorique}, postcode={91191}, city={Gif-sur-Yvette}, country={France}}
\affiliation[4]{organization={Universit\'e Paris-Saclay, CNRS, LPTMS}, postcode={91405}, city={Orsay}, country={France}}

\cortext[cor1]{Corresponding author}
\fntext[fn1]{These authors contributed equally to this work.}

\begin{abstract}
We reveal an interplay between temperature and radical concentration necessary to establish thermal mixing (TM) as an efficient dynamic nuclear polarization (DNP) mechanism. 
%investigate the temperature and radical concentration dependence of thermal mixing (TM) as a mechanism for dynamic nuclear polarization (DNP). 
We conducted DNP experiments by hyperpolarizing widely used DNP samples, i.e., sodium pyruvate-1-$^{13}$C in water/glycerol mixtures at varying nitroxide radical (TEMPOL) concentrations and microwave irradiation frequencies, measuring proton and carbon-13 spin temperatures. Using a cryogen consumption-free prototype-DNP apparatus, we could probe cryogenic temperatures between 1.5 and 6.5 K, i.e., below and above the {boiling} point of liquid helium. 
%For each probed condition, we determined the proton and carbon-13 spin temperatures.
%such that a "phase diagram" could be derived that illustrates the temperature and radical concentration dependence of TM efficiency. 
%The results reveal an interplay between temperature and radical concentration necessary to establish TM as an efficient DNP mechanism. 
We identify two mechanisms for the breakdown of TM: (i) Anderson type of quantum localization for low radical concentration, or (ii) quantum Zeno localization occurring at high temperature. 
%Not only do decreasing radical concentrations cause a breakdown of TM due to interruptions of the dipolar coupled network between the radicals, but quantum-Zeno effects also impede TM at higher temperatures. 
This observation allowed us to reconcile the recent diverging observations regarding the relevance of TM as a DNP mechanism by proposing a unifying picture and, consequently, to find a trade-off between radical concentration and electron relaxation times, which offers a pathway to improve experimental DNP performance based on TM.
\end{abstract}


\end{frontmatter}

\section{Introduction}

Dynamic nuclear polarization (DNP) is one of the most versatile techniques to enhance the intrinsically low signal-to-noise ratios of nuclear magnetic resonance (NMR) signals~\cite{abragam1978principles,wenckebach2016essentials}. 
It has become increasingly popular owing to its wide applicability in solid-state NMR through magic angle spinning (MAS)-DNP at temperatures close to 100 K~\cite{lesage2010surface,harrabi2022highly,alaniva2019electron,berruyer2022spatial}, and solution-state NMR~\cite{hilty2022hyperpolarized,otikovs2019natural,pinon2021hyperpolarization} or MRI~\cite{eichhorn2013hyperpolarization,nelson2013metabolic,comment2016dissolution} through dissolution DNP (DDNP) of ex-situ hyperpolarized samples below 2 K~\cite{ardenkjaer2003increase,van2016perspectives,kockenberger2007dissolution} followed by rapid sample heating and transfer to an MRI scanner or NMR spectrometer. 
In DNP experiments, a sample containing paramagnetic centers (most frequently stable nitroxide or trityl radicals)~\cite{kovtunov2018hyperpolarized,el2021sample} is submitted to a high magnetic field and irradiated with microwaves that excite electronic transitions. As a result, the high electron polarization is transferred to the  NMR-active nuclei surrounding the radicals, leading to an up to 10.000-fold~\cite{ardenkjaer2003increase} signal boost in comparison to conventional NMR at ambient temperatures. The often-encountered concentration and acquisition-time limitations of conventional NMR spectroscopy can thus be partially overcome.
Strikingly, despite the widespread use of DNP, the understanding of this non-equilibrium quantum process remains debated.\\
%Its efficiency relies on the polarization transfer from the unpaired electron spins of the radicals to the  nuclear spins. % due to electron-nuclear spin interactions.
In the past decades, three main mechanisms were established to explain the electron-nuclear polarization transfer in frozen solids: the solid effect (SE), the cross effect (CE), and thermal mixing (TM). 
For the well-resolved SE~\cite{abragam1978principles,wenckebach2016essentials,hovav2010theoretical}, microwave irradiation induces ``forbidden'' transitions in hyperfine-coupled two-spin systems composed of an electron and a nuclear spin, at microwave frequencies $\mw= \ome\pm\omn$, where $\ome$ denotes the electronic Zeeman and $\omn$ the nuclear Zeeman gap. 
The CE~\cite{hwang1967phenomenological,hovav2012theoretical} is {a model suitable} to explain DNP in samples containing bi-radicals with two populations of electron spins with Zeeman frequencies $\omega_{\rm e1}$, $\omega_{\rm e2}$. Under the resonance condition $\omega_{\rm e1}=\omega_{\rm e2}\pm \omn$, triple-spin-flip transitions transfer polarization between these electrons to nuclei, yielding nuclear polarizations of  $P_{\rm n}=\argp{P_{\rm e1}-P_{\rm e2}}/\argp{1-P_{\rm e1}P_{\rm e2}}$~\cite{hwang1967phenomenological,serra2013role,hovav2015effects}, with $P_{\rm e1}$, $P_{\rm e2}$ the electronic polarizations. These transitions rely on electron-electron dipolar and electron-nuclear hyperfine interactions that induce mixing between the quantum states of the three spin species. For mono-radicals the CE is also important when the width of the electron paramagnetic resonance (EPR) line is larger than $\omn$~\cite{hovav2015effects}. \\
However, when the concentration of unpaired electrons becomes large enough, three-spin mixing events alone do not suffice to account for the hyperpolarization process. Instead, many-spin transitions must be considered, which lead to a coupling of the entire ``bath'' of electrons to the nuclei. {In such a situation, a so-called thermal mixing (TM) of the spin species is assumed. This concept is based on the fact that the off-equilibrium quantum steady state of the electron spins is well described by a} \textit{thermodynamic density matrix with two temperatures}~\cite{provotorov1962magnetic,borghini1968spin,abragam1978principles,de2015dynamic,maimbourg2021bath,jahnig2019spin}: {the spin temperature $T_{\rm s}$ of the non-Zeeman reservoir and $T_{\rm Z}$ the temperature of the Zeeman reservoir. }
The nuclear species with Zeeman frequency below the non-Zeeman electronic energy scale will freeze at the same spin temperature as the electrons, $T_{\rm s}$. The condition for strong nuclear hyperpolarization is then $T_{\rm s}\ll T\ll T_{\rm Z}$. \\
However, the experimental relevance of TM is still controversial. 
Its main fingerprint is the observation of a common spin temperature $T_{\rm s}$
for all the nuclear species of the sample (${}^1$H, ${}^{13}$C, ${}^{14}$N, ...).~\cite{kurdzesau2008dynamic,lumata2011dnp}. 
A second direct evidence of TM was demonstrated a long time ago by Atsarkin~\cite{atsarkin1970verification} on ${\rm Ce}^{3+}$ half-spin impurities in a ${\rm CaWO}_4$ crystal devoid of nuclear spins (so that three-spin mixing processes are absent). He observed that the entire EPR profile is modified by microwave irradiation (including frequencies $\om\neq\mw$). The spectrum even partially displayed an unusual reversal of polarization, an observation well compatible with a TM description by the two temperatures $T_{\rm s}$, $T_{\rm Z}$. Nonetheless, recent electron double resonance (ELDOR) experiments performed at various temperatures and radical concentrations challenged the relevance of TM~\cite{hovav2015electron,kaminker2016heteronuclear,kundu2018electron,kundu2019experimental} for experimental DNP. In particular, a large reorganization of the EPR spectrum was observed under irradiation but without any reversal of polarization. {These measurements were explained not with TM but with a model combining a phenomenological spectral diffusion}\cite{hovav2015electron,kundu2018electron,kundu2019experimental}  (see also\cite{C6CP04345C}) {and the CE, called \textit{direct} CE when one of the electron spins involved in the three-spin mixing is irradiated, and \textit{indirect} CE when both electron spins are not irradiated. Another resonance called \textit{heteronuclear} CE, involving two electron and two nuclear spins, was also considered in} ~\cite{kaminker2016heteronuclear}. 

Herein, we aim to resolve this discrepancy between these new results suggesting the absence of TM and the older direct observations of TM. 
For this purpose, we performed spin-temperature measurements under DNP of ${}^1$H and ${}^{13}$C nuclear spins varying both radical concentration and temperature. We observe a breakdown of TM under two conditions: either at high temperature ($T\gtrsim 5$-$10$ K) or at low radical concentrations. Consequently, TM is not a relevant mechanism for MAS-DNP that occurs around $100$ K. It is instead 
the dominant regime for dissolution DNP (around $2$ K) and controls the performance of the experiment. 
We rationalize these observations as a competition between the electron relaxation time and the time associated with spectral diffusion. Combining the present results together with earlier reports~\cite{guarin2017characterizing}, we propose a phase diagram that conciliates in a unified picture the above-mentioned experiments and indicates where efficient nuclear hyperpolarization by TM is expected. 

\section{Experimental section}
DNP experiments were performed on a home-built cryogen consumption-free prototype based on a Cryogenic LTD cryostat-magnet combination described in Ref.~\cite{kress2021novel}. In brief, $1.5$ M sodium pyruvate-1-$^{13}$C were dissolved in freshly prepared solutions of varying TEMPOL (4-Hydroxy-2,2,6,6- tetramethylpiperidinyloxyl) concentrations (40, 60, and 70 mM) in 50\% v/v glycerol-d$_8$, 40\% D$_2$O, and 10\% H$_2$O. 100 µL of each sample were then hyperpolarized for $1.5$ h at temperatures of $1.5$, $3.5$, $5$ or $6.5$ K, in a magnetic field of $B_0 = 6.7$ T employing $50$ mW microwave irradiation. We used a Virginia Diodes microwave source coupled to a 16-fold frequency multiplier. For microwave sweep experiments the irradiation frequency was varied between $187.5$ and $188.5$ GHz. The maximum DNP efficiency for spin-temperature experiments was at $188.0$ GHz. NMR signals were recorded every 5 s using 5° excitation angles obtained with radio frequency pulses at 285.3 MHz for $^1$H or 71.3 MHz for $^{13}$C, produced by a customized Bruker HDIII spectrometer console.

{All samples were degassed by purging with nitrogen gas under vacuum for 30 min.}

For data analysis, the recorded signals were baseline corrected using polynomial fit functions and apodized using a Gaussian window function. The signals after Fourier transformation were then fitted to derive the signal amplitudes $S$. The resulting values were then extrapolated to infinite times using mono exponential growth functions.

\begin{figure}[h!]
    \centering
   \includegraphics[clip, trim=6cm 3cm 6cm 3cm, width=1.00\textwidth]{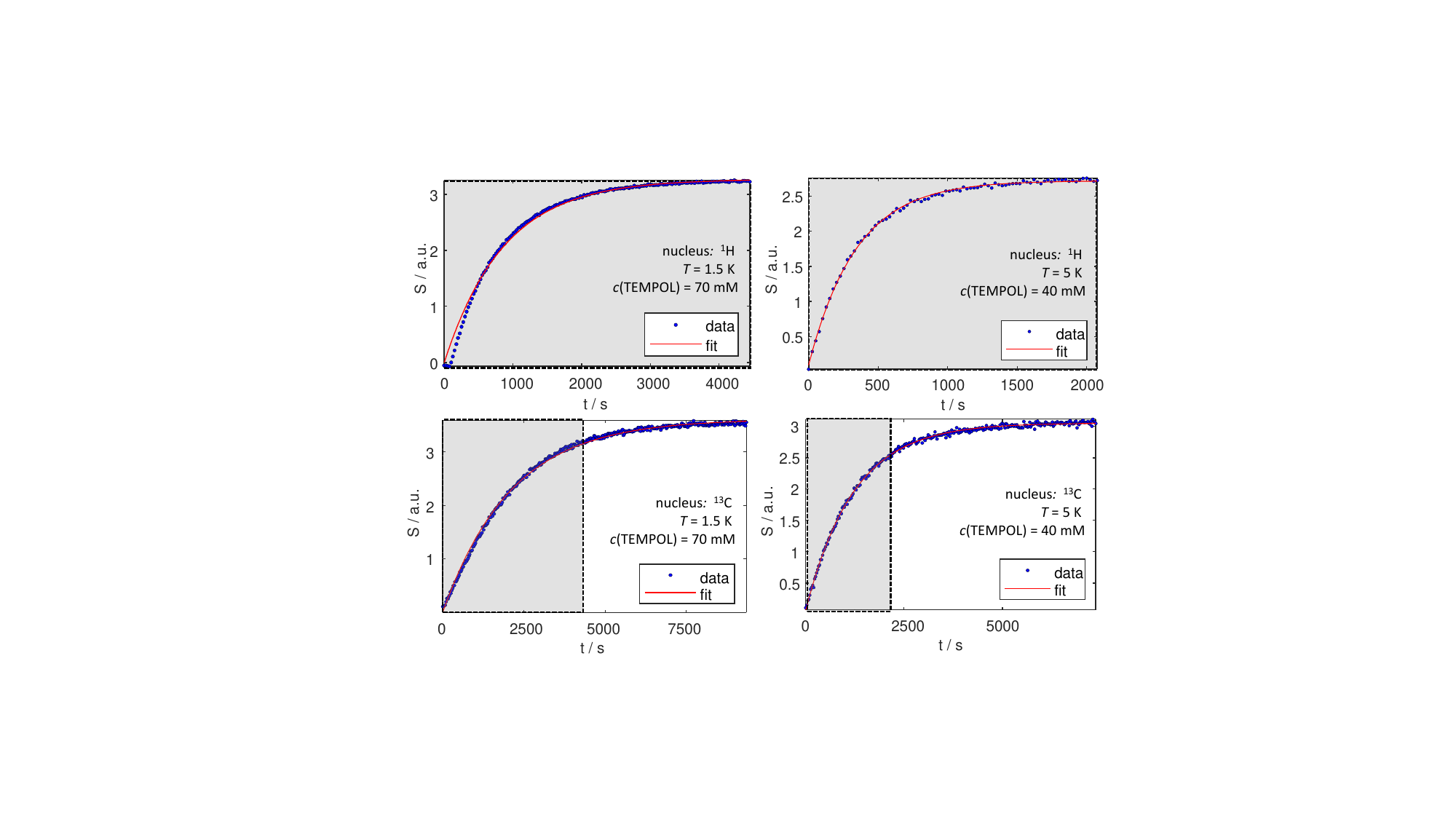}
    \caption{Signal build-up functions for two limiting cases: low radical concentrations and high temperature vs. high radical concentration and low temperature. In both cases, the build-up rates are different for proton and carbon-13 spins. The grey shades indicate equivalent periods of time.}
   \label{fig:budata}
\end{figure}

Exemplary build-up curves and fits are shown in Fig.~\ref{fig:budata}. The extrapolated signal amplitudes at infinite times correspond to steady-state signal amplitudes. Hence, varying build-up rates did not influence our analysis.

\subsection{Methodological Considerations}
{All 260 recorded signal build-ups, the corresponding fits, and the build-up rate constants are shown in the Supporting Information. Given the high signal-to-noise of the DNP experiments, the errors in the fits varied between 1\% and 5\%. Repetitions of entire experiments (including sample preparation) lead to errors in the spin temperatures of +/-0.0004 K to +/-0.003 K.

Similarly, the DNP profiles and build-ups were well reproducible. Repeating selected measurement series led to excellent agreement with the original experiments (data again shown in the Supporting Information). The major source of error in the reported experiments is likely  the sample temperature. All reported temperatures are those of the helium bath. Yet, the sample temperature might differ by up to +/-0.3 K due to microwave-induced heating and unknown temperature conduction between the variable temperature insert and the sample.}

{The used DNP prototype system operates in a cryogen consumption-free mode, which allows to record signals in thermal equilibrium even despite very long longitudinal relaxation times (see references} \cite{negroni2022fremy,kress2021novel} {for a description of the system). All build-up curves are shown in the Supporting Information. The absolute signal intensities for a given temperature and frequency did not vary strongly upon changing the radical concentrations. However, it should be noted that for the determination of the spin temperature, comparisons of signal intensities between thermal equilibrium and hyperpolarized cases are relevant only for a given sample.

The capacity to record very long time series also allowed us to record reliable background signals by measuring the signal of an empty sample cup (see the Supporting Information) and subtracting these from the data recorded under DNP conditions.}

\section{Results}
The signal amplitude $S$ under irradiation is proportional to the nuclear polarization $P_{\rm n}$. Hence, to determine the spin temperatures $T_{\rm s}$ through $P_{\rm n}$, one needs to determine the proportionality constant in $S\propto P_{\rm n}$. This is done by measuring the signal under off-resonance microwave irradiation of 187.52 GHz, i.e., in the absence of any DNP effect. There, the signal corresponds to the thermal equilibrium polarization $S_{\rm eq}\propto P_{\rm n,eq}$. As $P_{\rm n,eq}$ is well known theoretically, $P_{\rm n}$ can be derived for generic irradiation frequency. For the spin temperature one then finds:

\begin{equation}\label{eq:Pn_33}
 T_{\rm s}=\frac{\hbar\omega_{\rm  n}}{2k_B\arctanh(P_{\rm n})}
\end{equation}

%To experimentally probe this duality of TM breakdown, we measured the spin temperatures TS of proton and carbon nuclei under DNP at varying sample temperatures (1.8, 3.5 and 5 K) and radical concentrations (40, 60, and 70 mM) using the nitroxide TEMPOL. The results are shown in Fig. 2. It can be seen that TS of both spin types are similar at high TEMPOL concentrations of 70 mM, independent of the sample temperature. Contrarily, at lower concentrations of 40 mM, the TS differ at any temperature. At intermediate TEMPOL concentration of 60 mM, TS are similar only a temperatures below 4 K.

\begin{figure}
    \centering
   \includegraphics[clip, trim=0cm 11cm 0cm 11cm, width=1.00\textwidth]{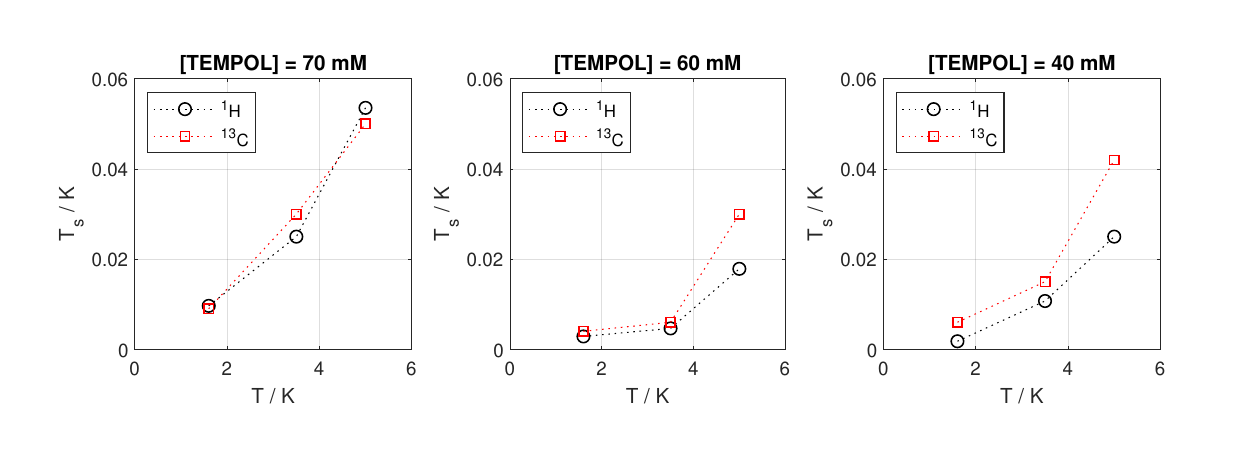}
    \caption{Spin temperatures $T_{\rm S}$ with varying TEMPOL concentrations and sample temperatures $T$  for $^1$H and $^{13}$C nuclei within the same sample. 
    % With decreasing temperature and increasing concentrations, the spin temperatures tend to equalize.
    The points for $70$ mM at $6.5$ K are not reported because $P_{n,eq}$ is too small with respect to the noise.}
   \label{fig:expdata}
\end{figure}

The spin temperatures resulting from this procedure are shown in Fig.~\ref{fig:expdata} for the herein-probed radical concentrations, temperatures, and nuclei, i.e., $^1$H and $^{13}$C. At high radical concentrations of 70 mM, both spin temperatures are compatible between 1.5 K and 5 K for both nuclei. In contrast, at lower concentrations of 40 mM TEMPOL, the spin temperatures never coincide. At 60 mM TEMPOL, the spin temperatures merge only below a temperature of 3.5 K. We interpret differing $T_{\rm s}$ values as a breakdown of TM. Hence, at high enough temperatures or low radical concentrations, the DNP process does not appear to be dominated by TM.

{It should be noted that the spin temperatures one can obtain depend strongly on the sample composition and the used DNP system. Protonation degree, analyte, and radical concentration, as well as microwave irradiation can significantly impact the performance of a DNP protocol, e.g., by changing spin and spectral diffusion kinetics.}\cite{leavesley2017effect} {For example, J\"ahnig et al. corroborate the herein-reported interpretation of a unique spin temperature by demonstrating magnetic field dependence of thermal mixing. Yet, they report different polarization values at similar temperatures as fully protonated samples were used, which changes the point when the DNP mechanism switches from thermal mixing to other processes. Therefore, spin temperatures and polarization should only be quantitatively compared for a given sample.}\cite{jahnig2019spin} 
{Therefore, spin temperatures and polarization should only be quantitatively compared for a given sample.}

To further corroborate our interpretation and analyze the data obtained (especially at $70$ mM) in more detail, we make use of the fact that the spin temperature is the only parameter in the hyperpolarization $P_n$  that depends on the microwave frequency and that, when the spin temperature is not too small compared to the nucleus Zeeman gap, the relation between $P_n$ and the spin temperature $T_S$ becomes linear:
\begin{equation}\label{eq:Pn_3}
P_{\rm n}=\tanh{\frac{\hbar\omega_{\rm  n}}{2k_B T_{\rm s}}} \sim \frac{\hbar\omega_{\rm  n}}{2k_B T_{\rm s}}
 %T_{\rm s}=\frac{\hbar\omega_{\rm  n}}{2k_B\arctanh(P_{\rm n})}
\end{equation}
Therefore, when the spin temperatures of both species match, the signals as a function of the microwave frequency must display an identical shape.

\begin{figure}[h!]
    \centering
    \includegraphics[clip, trim=8cm 0cm 9cm 0cm, width=1.0\textwidth]{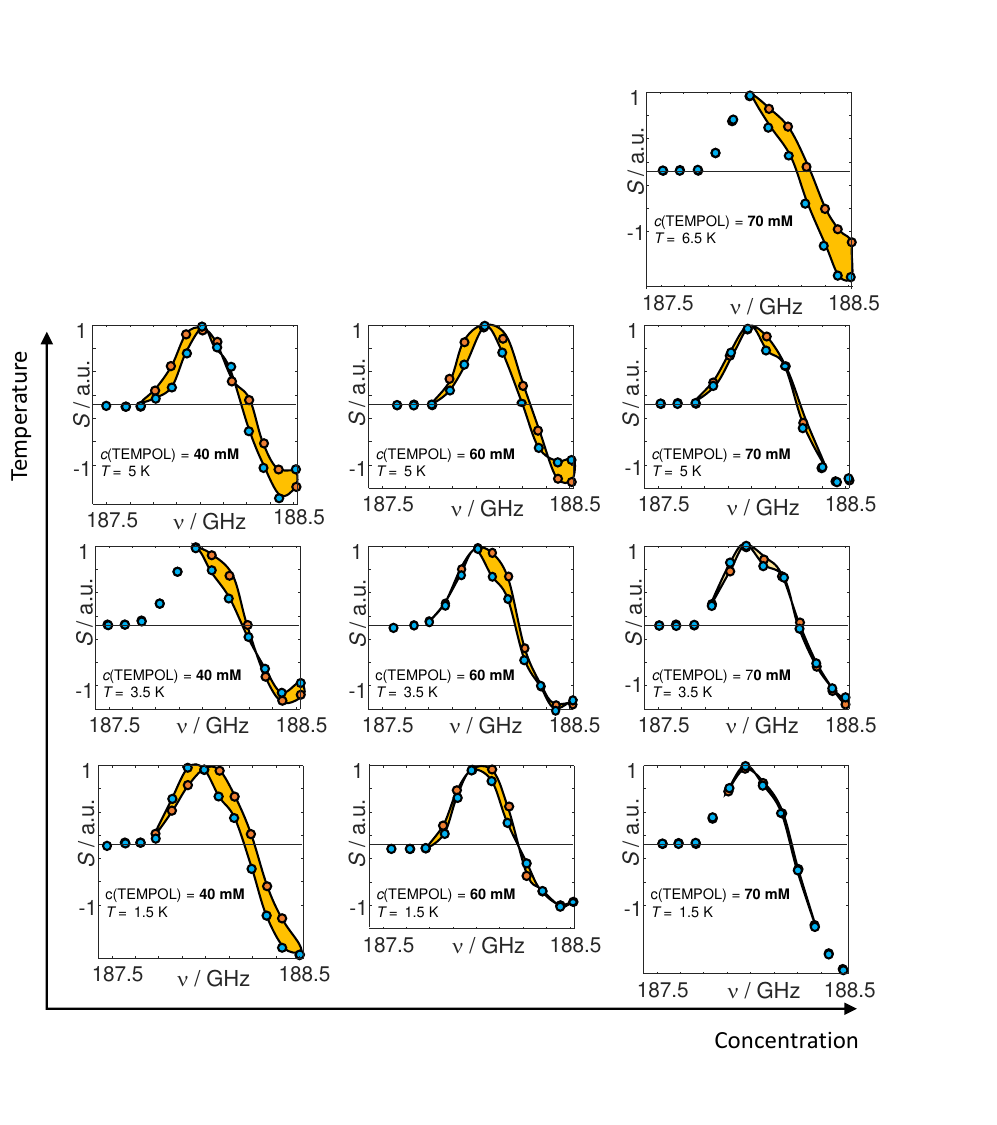}
    \caption{Signal intensities $S$ in dependence of the microwave irradiation frequency $\nu$ for $^1$H (blue) and $^{13}$C (orange) spins for varying concentrations and temperatures. The yellow shades highlight the differences between proton and carbon profiles. {The frequency domain was corrected for B$_0$ fluctuations, such that the zero-crossings conincide. The original data is shown in the Supporting Information.}}
    \label{fig:expdata2}
\end{figure}

In Fig.~\ref{fig:expdata2}, we show the microwave sweep for each probed condition. We normalized the $^1H$ and $^{13}C$ profiles to their respective maxima. If thermal mixing holds, the obtained curves need to be identical, as both profiles rescale each other according to Eq.~\eqref{eq:Pn_3}. Hence, we interpret a difference in these curves as a breakdown of TM.
This method has the advantage that no thermal equilibrium signal intensity is required for referencing. Consequently, we manage to probe the breakdown of TM at TEMPOL concentrations of 70 mM at temperatures above $5$ K, where the non-hyperpolarized signal intensities become prohibitively weak with our prototype setup.

At high TEMPOL concentrations of 70 mM, TM is present only below $6.5$ K; at $60$ mM, it starts to emerge below $5$ K. Instead, for $40$ mM, we have no sign of TM at any of the probed temperatures. 
Note that the reported signal amplitudes again correspond to infinite build-up times. Interestingly, though, the build-up rate constants for $^1H$ and $^{13}C$ differ for all conditions probed herein (Fig.~\ref{fig:budata}), even when spin temperatures converge at long times. In the words of Wenckebach\cite{wenckebach2016essentials}, the probed TEMPOL-based radical system thus appears to transition into a regime of ``slow'' TM at increasing concentrations and decreasing temperatures.

%It should be noted that the $T_{\rm S}$ is likely to diverge at temperatures >6.5 K for the case of 70 mM TEMPOL. However, with the used experimental setup, such conditions could not be probed in a feasible manner.
%The contribution of TM to DNP in our experiments is furthermore reflected in the dependence of the relative NMR signal intensities on the microwave irradiation frequency. In the case of TM, this dependency is similar for all nuclear spins in the sample, i.e., if only one common spin temperature is adopted. Fig.~\ref{fig:expdata2} exemplary shows comparisons of microwave sweep profiles for $^1H$ and $^{13}C$ spins at low (40 mM) to high (70 mM) radical concentrations at varying temperatures. With increasing concentration and decreasing temperatures, the profiles for the two nuclei become increasingly similar, hence, indicating the dominance of TM as DNP mechanism. This finding again supports our conclusion that higher radical concentrations and lower temperatures favor DNP by TM. 

\section{Discussion}

\begin{figure}[h!]
 \centering
 \includegraphics[clip, trim=3cm 4cm 3cm 5cm, width=1.0\textwidth]
 {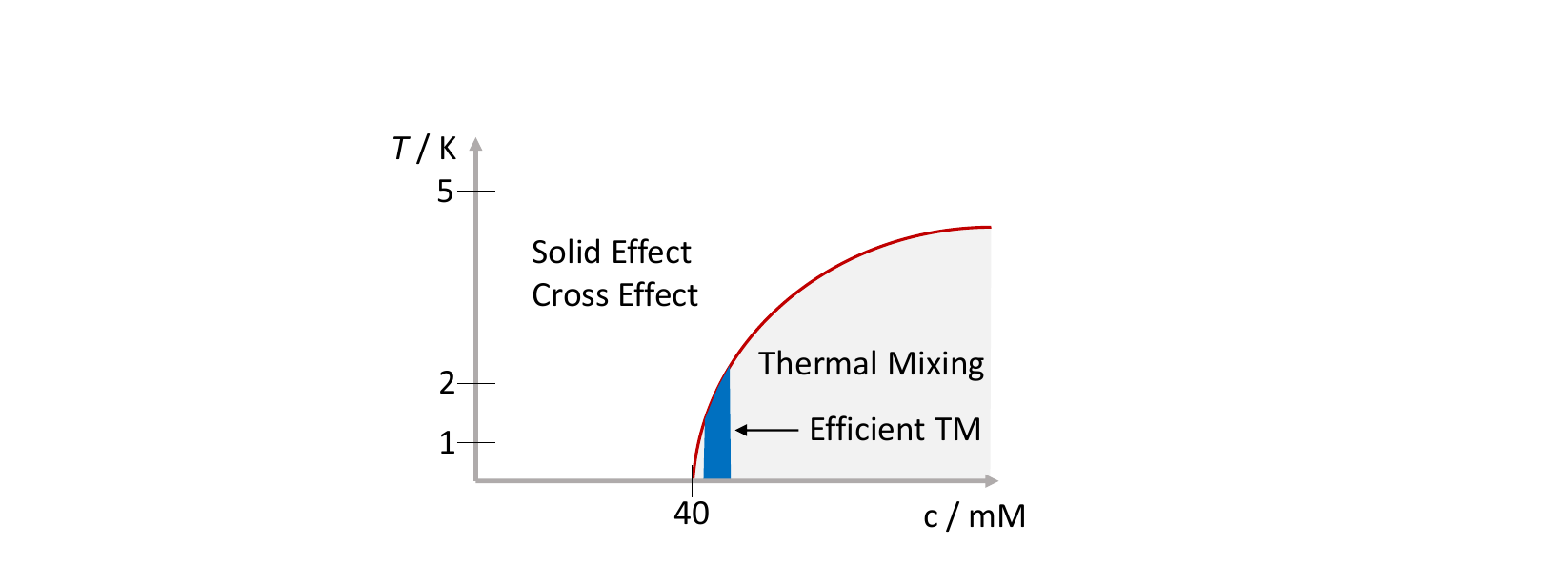}
     \caption{Phase diagram relating lattice temperature $T$ and the radical concentration with different DNP regimes. For the samples probed herein, a maximum DNP efficiency was observed at the verge of TM breakdown, i.e., close to 40 mM.}
 \label{fig:phasediagram}
\end{figure}

We propose a summary of our investigations in the diagram in Fig.~\ref{fig:phasediagram} and discuss the hyperpolarization behavior and establishment of TM below.

\begin{figure}
    \centering
    \includegraphics{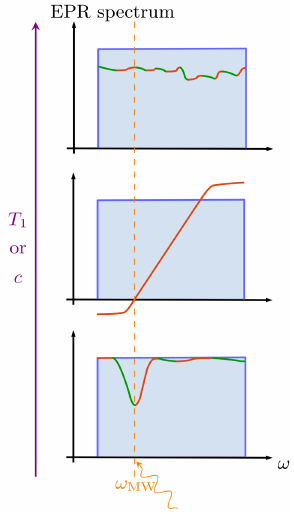}
    \caption{Sketches of the EPR spectra under irradiation at frequency $\omega_{\rm MW}$ assuming varying electron relaxation times $T_1$ or radical concentrations $c$ showing the cases of homogeneous TM (top), inhomogeneous TM (center), and TM breakdown (bottom). Red ({respective} green) indicates frequency intervals in which the numerator's integrand of Eq.~\eqref{eq:Pn} is negative ({respective} positive). The blue line represents the equilibrium EPR line without any microwave irradiation.}
    \label{fig:sketch}
\end{figure}

The performance of a DNP protocol for hyperpolarizing nuclear spins
%in the various physical situations considered here (according to radical concentration, temperature, Gadolinium doping) 
can be quantified through the following formula for the nuclear polarization~\cite{serra2013role} 
\begin{equation}\label{eq:Pn}
 P_{\rm n}=\frac{\int \dd\om\,[P_{\rm e}(\om)-P_{\rm e}(\om+\om_{\rm n})]}{\int \dd\om\, [1-P_{\rm e}(\om)P_{\rm e}(\om+\om_{\rm n})]}
\end{equation}
where $P_{\rm e}(\om)$ is the electronic polarization at frequency $\om$. This formula assumes an efficient polarization transfer through triple spin flips yet neglects nuclear relaxation processes with the lattice, as these are typically very slow ($> 2 $ h at $1.4$ K)~\cite{negroni2022fremy}. For simplicity, we assumed a rectangular EPR line for Eq.~\eqref{eq:Pn}, as displayed in Fig.~\ref{fig:sketch}. However, it can be readily generalized to arbitrary line shapes. Under microwave irradiation, the EPR spectrum is rearranged such that nuclear hyperpolarization arises from the imbalance of the electron polarizations $P_{\rm e}(\om)-P_{\rm e}(\om+\om_{\rm n})$ in the numerator of Eq.~\eqref{eq:Pn}.  
Three situations can occur: \textit{(i)} Fig.~\ref{fig:sketch} (top): homogeneous TM. When all electron spins are very rapidly mixed, the irradiated EPR line is uniformly reduced. As a result, the sample is poorly hyperpolarized. 
\textit{(ii)} Fig.~\ref{fig:sketch} (middle): inhomogeneous TM. In this regime, the electron spins are still mixed, yet the irradiated EPR line is asymmetric and displays the partial reversal of polarization as often observed in DNP applications. This is the best regime for nuclear hyperpolarization, as the imbalance of the electron polarizations $P_{\rm e}(\om)-P_{\rm e}(\om+\om_{\rm n})$ accumulates.  
\textit{(iii)} Fig.~\ref{fig:sketch} (bottom): TM breaks down. Here the electron spins are not mixed. A spectral hole is ``burnt'' close to irradiation frequency. The imbalance changes sign across the irradiation frequency, implying a cancellation leading to weak or even negligible hyperpolarization. 

{Two control parameters allow one to switch from one regime to the other: the radical concentration $c$ and the longitudinal electron relaxation time $T_1$ (itself temperature dependent}\cite{C4CP02636E,C6CP00914J}{, decreasing as temperature raises). These two handles have recently been established independently by Guarin et al}.~\cite{guarin2017characterizing} and Maimbourg et al~\cite{maimbourg2021bath}.
{Taken together, these works suggest that high radical concentration or long relaxation times lead to homogeneous TM. On the contrary, if the radical concentration is too low or relaxation too fast, TM breaks down}~\cite{jahnig2019spin,hovav2015electron,kaminker2016heteronuclear,kundu2018electron,kundu2019experimental}. {Combining the information from all these studies with the temperature dependence reported herein, a unified description of DNP comes into reach.}

As a consequence, the {recipe} for a high nuclear hyperpolarization relies on the trade-off between radical concentration and relaxation time to induce inhomogeneous TM. Interestingly, at a given low temperature, the best performance has been observed at the verge of the TM breakdown in terms of radical concentration~\cite{guarin2017characterizing}. 

TM breakdown is a novel aspect of comprehending nuclear hyperpolarization at cryogenic temperatures. %It takes place under a variation of the two control parameters and
{In the DNP community, the presence of a single spin temperature is often traced back to the efficiency of spectral diffusion. Our results show that there are two processes competing against spectral diffusion. The first is the g-anisotropy that hinders the spectral diffusion. The second occurs when the single spin-flip transitions induced by the lattice are fast with respect to spectral diffusion. 
Interestingly, in many-body physics, the efficiency of the spectral diffusion is interpreted as a competition between ergodicity and localization. Also, in this context, the origin of localization is twofold:}

{(i) decreasing the radical concentration and, thus, the strength of dipolar interactions against the disorder imposed by the g-factor anisotropy. This mechanism is due to Anderson localization}~\cite{anderson1958absence} {and its many-body generalization}~\cite{abanin2019colloquium}{: below a critical ratio between interaction and disorder strengths, the eigenstates of the electron spin Hamiltonian become non-ergodic and prevent the quantum dynamics of the electron spins to reach a thermodynamic state. In the DNP context, it has been analyzed in detail in Refs}.~\cite{de2015dynamic,RAMRDL18}.

{(ii) increasing the temperature and, thus, the rate of the transitions induced by the lattice. During these single spin flips, local operators such as $S^+_i, S^-_i$, $S_i^z$ act as measurements on the quantum state of the electron spins. When the rate of such quantum measurements becomes large enough, the quantum dynamics of the spectral diffusion is frozen, as studied in detail in Ref}.~\cite{maimbourg2021bath}{. This second localization mechanism is akin to the quantum Zeno effect}~\cite{misra1977zeno,facchi2001quantum}{, in which the frequency of measurements is high enough that the state of the electron spins cannot decay into eigenstates of the electron spin Hamiltonian. In other words, the steady-state density matrix is no longer diagonal in the eigenstates of the electron spin Hamiltonian. In this case, these eigenstates are truly ergodic, but the high rate of single spin flips precludes a thermodynamic description of the steady state.}

\section{Conclusions}

In conclusion, we have discovered that the efficiency of spectral diffusion, which underlies the effectiveness of TM in cases of heterogeneously broadened electron paramagnetic resonance (EPR) lines, depends not only on the radical concentration but also on the experimental temperature. 
Assuming converging spin temperatures indicate the presence of thermal mixing (TM), our experimental data demonstrate that low sample temperatures and high enough radical concentrations favor TM as the predominant dynamic nuclear polarization (DNP) mechanism. % This alignment between our experimental observations and the theoretical predictions mentioned earlier is summed up in Fig.~\ref{fig:phasediagram}, which presents a phase diagram illustrating the emergence of TM as a function of temperature and radical concentration.
Regarding the discrepancy between studies supporting TM as an efficient DNP mechanism and contradictory reports, our theory and data shed new light on this debate. The underlying studies have been conducted at different temperatures ranging from $1.2$ to $20$ K. It is thus highly likely that the diverging observations reported can be attributed to variations in the effectiveness of the quantum Zeno effect at different temperatures.

While it is intuitive to assume that increasing concentrations enhance the occurrence of electron flip-flop events that distribute information throughout the EPR spectrum, the temperature dependence is less apparent. However, considering that higher temperatures (and the resulting increased relaxation rates) lead to the occurrence of a quantum Zeno effect that freezes spectral diffusion processes, these dual conditions of TM efficiency become theoretically predictable and observed in our experiments. 
Our data demonstrate that an optimal hyperpolarization based on TM can be achieved at the crossover between TM and other mechanisms, such as the solid effect. This trade-off between radical concentration and electron relaxation times provides valuable insights into the controversy surrounding the predominance of TM as a DNP mechanism and may enhance existing DNP approaches to maximize nuclear polarization.

\section{Conflicts of Interest}
The authors declare no conflicts of interest.

 \section{Acknowledgements}
 
The authors thank the NMR core facility of the University of Vienna. We also thank Quentin M. Riedl, David O. Guarin, Denis M. Basko and Markus Holzmann. The project received funding from the European Research Council (ERC) under the European Union's Horizon 2020 research and innovation program (grant agreement 801936). 

 \section{Supporting Information}
 Data reproduction, build-up curves, background correction.

 \section{Data Availability}
The raw data is available under https://phaidra.univie.ac.at/o:2058312.

\bibliographystyle{elsarticle-num-names}
\bibliography{DNP.bib}

 

\section*{Supplementary material}

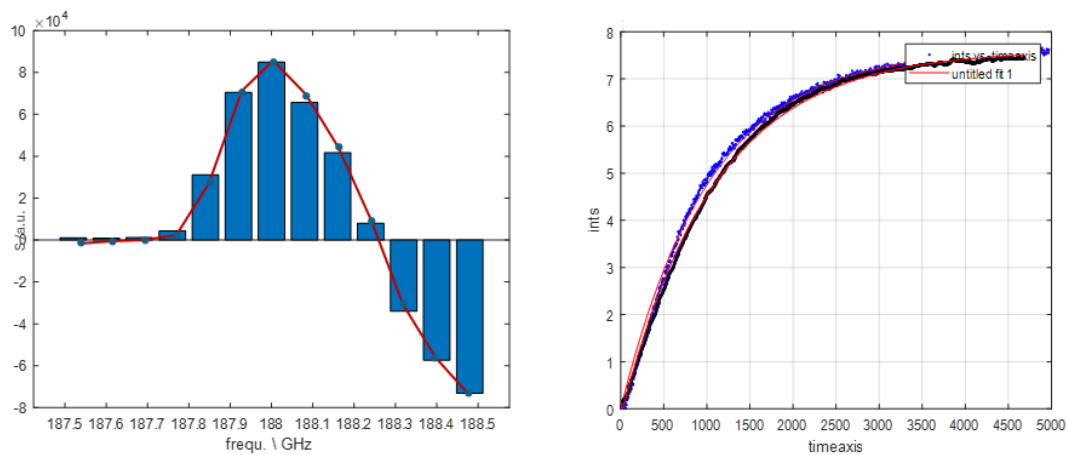

**Figure S1. Two examples of reproducibility. Left:** Exemplary data reproduction for c(TEMPOL) = 70 mM, T = 3.5 K, nuc: $^{13}$C. Blue bars are the original experiment; the red curve is a reproduction. Right: Original (blue) and reproduced $^{13}$C build-up (black) at 40 mM TEMPOL and 1.5 K.

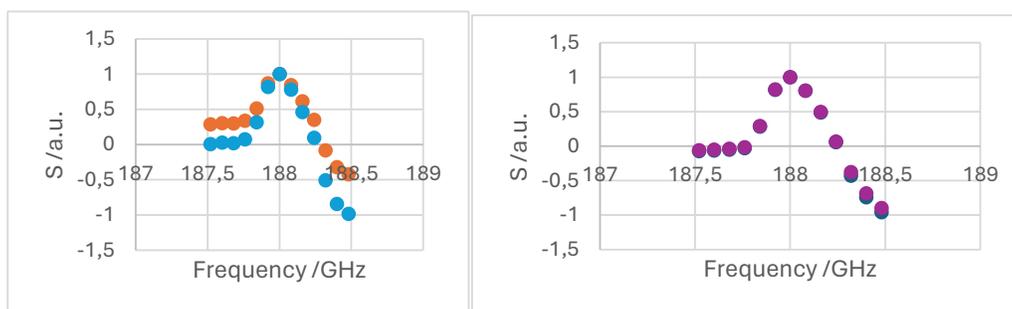

**Figure S2.** Normalized $^1$H (left) and $^{13}$C (right) DNP profiles with (blue and violet) and without (orange and green) background correction at 3.5 K and 40 mM TEMPOL. Similar corrections have been used for all experiments.

**Build-up curves and times**

Fig. S4-S6 shows that the extrapolation procedure is working well, and we can determine the signal intensity under steady-state conditions. The effect of different build-up times, *i.e.*, slower build-up at lower temperatures, lower radical concentrations, and when switching from $^1$H to $^{13}$C, is considered.

**Figure S3.** Signal build ups (blue) and fitted exponentials (red) for a concentration of 40 mM TEMPOL. Temperatures and nuclei are indicated in the figure.

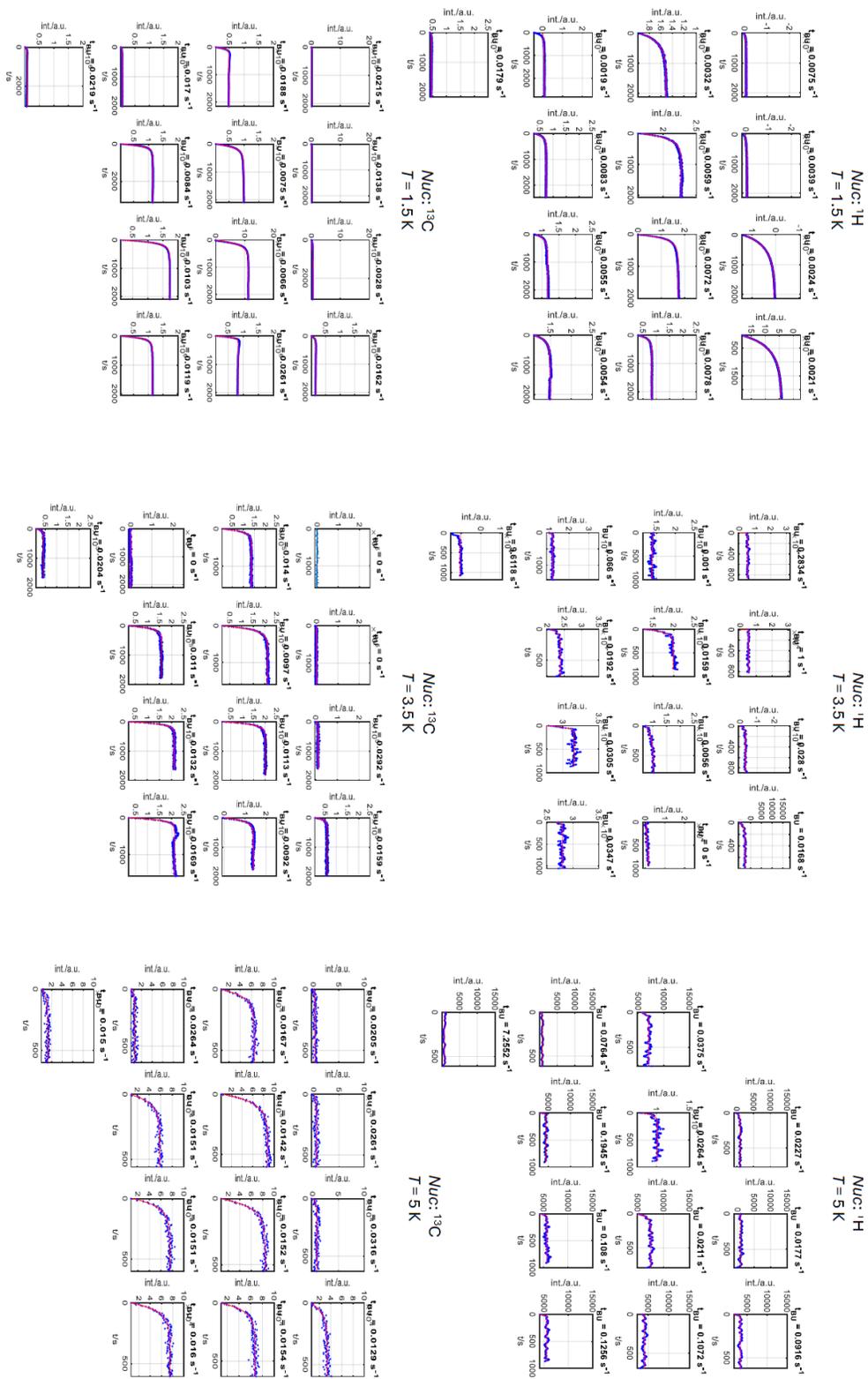

**Figure S4.** Signal build-ups (blue) and fitted exponentials (red) for a concentration of 60 mM TEMPOL. Temperatures and nuclei are indicated in the figure.

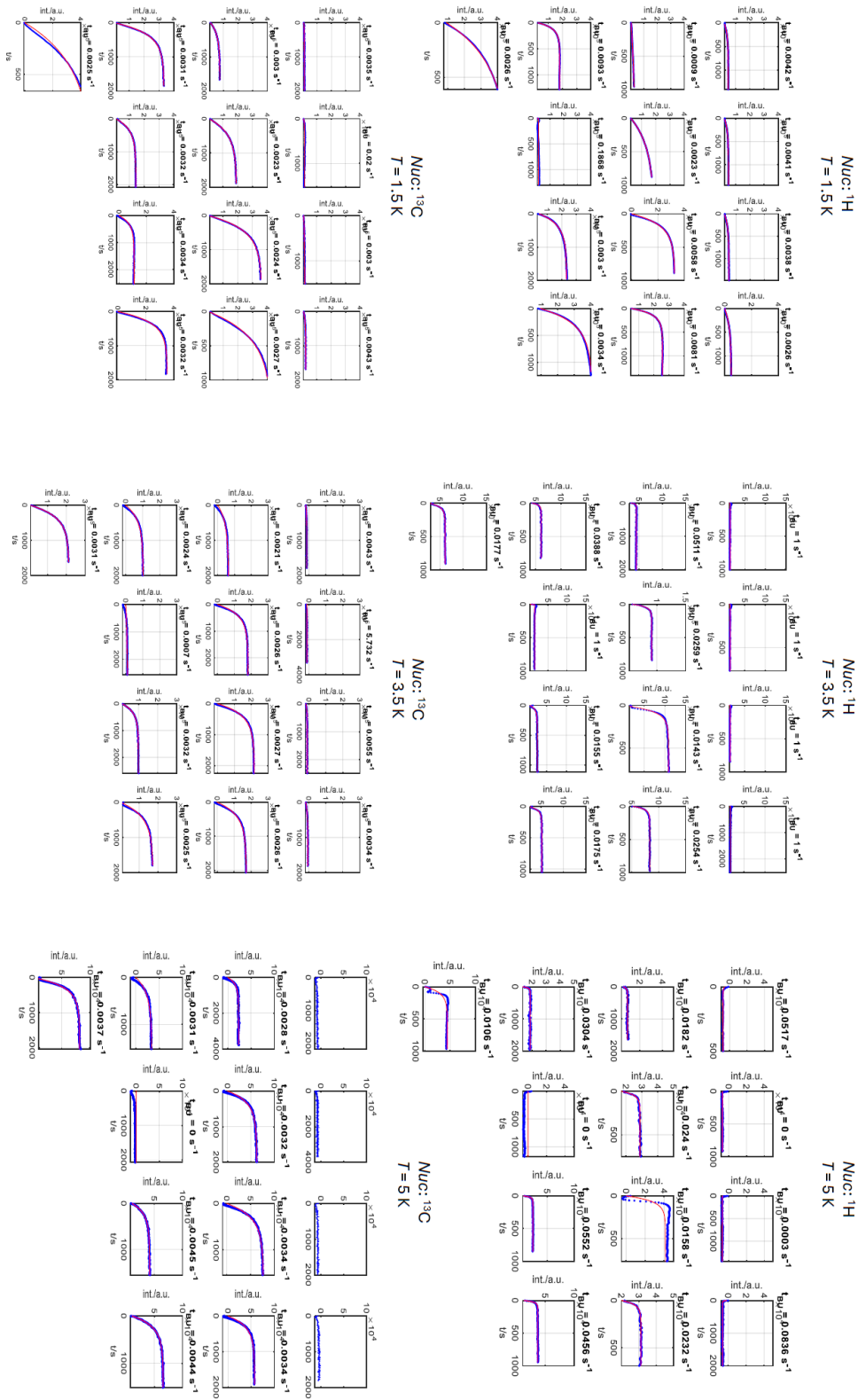

**Figure S5.** Signal build ups (blue) and fitted exponentials (red) for a concentration of 70 mM TEMPOL. Temperatures and nuclei are indicated in the figure. It should be noted that the build-ups are so fast in

a few cases, such that the build-up time was not correctly fitted by our MATLAB routine (showing values of 0 or >= 1 s-1). Yet, the extrapolated intensities at infinite times are not affected by an artificially fast build-up.

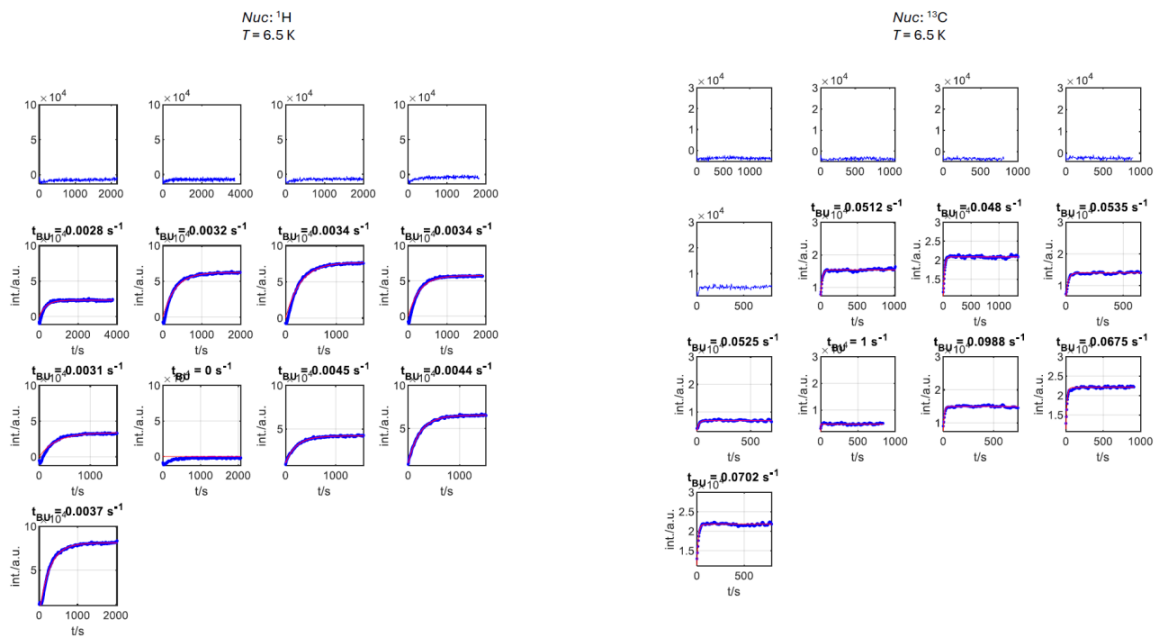

**Figure S6.** Signal build-ups (blue) and fitted exponentials (red) for a concentration of 70 mM TEMPOL. Temperatures and nuclei are indicated in the figure. Missing build-up times indicate that the intensities remained too close to the noise level for reliable fitting.

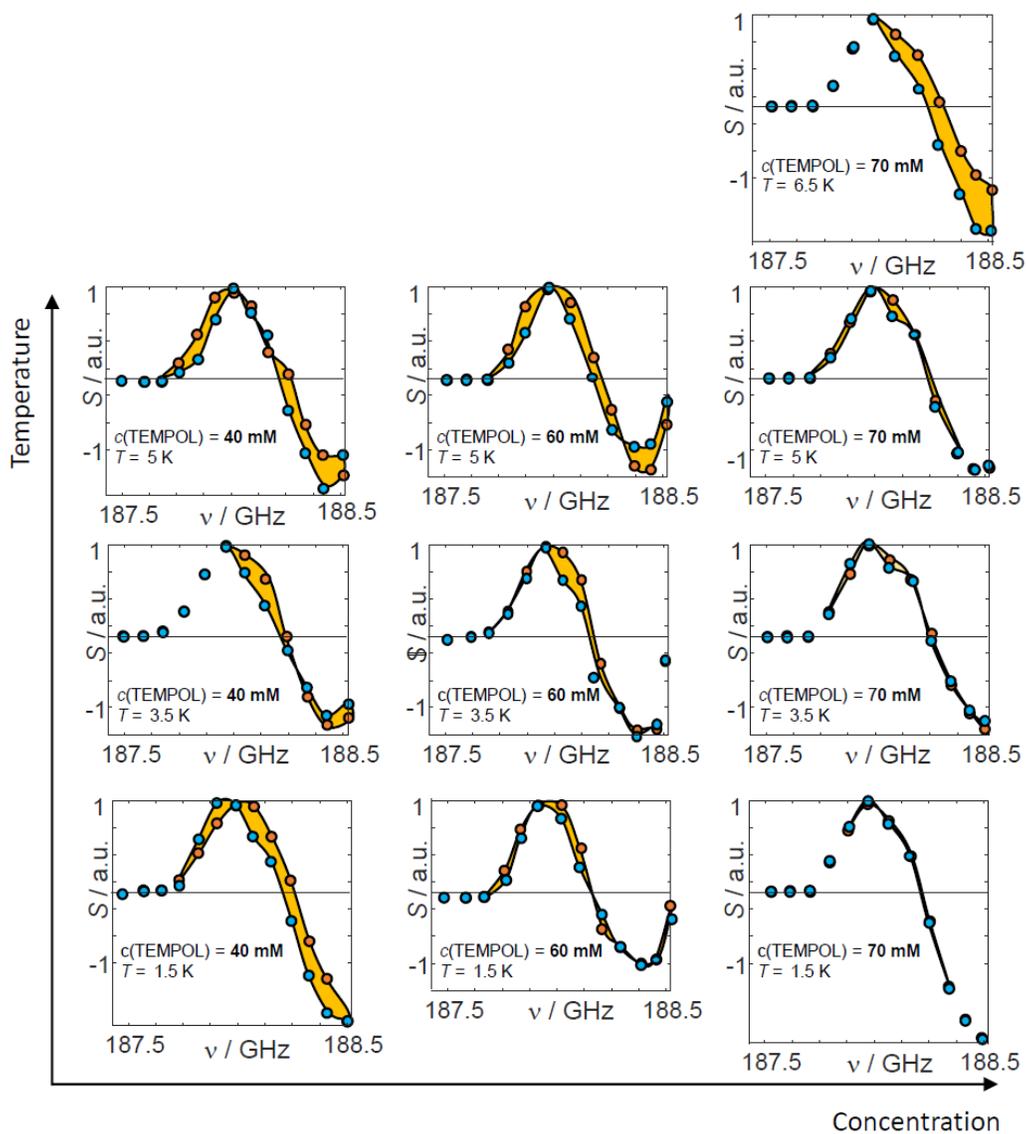

**Figure S7.** Original, not frequency-corrected data underlying Fig. 3 of the main text. The zero-crossings of the profiles do not all coincide, as ramping the magnet causes changes in the exact $B_0$ field, which translates into different frequency dependencies of the efficiency of microwave irradiation. Thus, the 60 mM profile is shifted to lower frequencies. However, it is important to note that the overall effect on the DNP profile is clear in all three samples: decreasing the temperature leads to the emergence of thermal mixing.

 

\end{document}